%% file: new-proof-style-arxiv.tex
\documentclass{article}

\usepackage{amssymb}
\usepackage{wasysym}
\usepackage{url}
\usepackage{graphicx}

\newcommand{\be}{\begin{enumerate}}
\newcommand{\ee}{\end{enumerate}}
\newcommand{\bi}{\begin{itemize}}
\newcommand{\ei}{\end{itemize}}
\newcommand{\bc}{\begin{center}}
\newcommand{\ec}{\end{center}}
\newcommand{\bsp}{\begin{sloppypar}}
\newcommand{\esp}{\end{sloppypar}}
\newcommand{\high}{\rotatebox[origin=c]{90}{\CIRCLE}}
\newcommand{\medhigh}{\rotatebox[origin=c]{90}{\RIGHTcircle}}
\newcommand{\medlow}{\rotatebox[origin=c]{-90}{\RIGHTcircle}}
\newcommand{\low}{\rotatebox[origin=c]{90}{\Circle}}
\newcommand{\mname}[1]{\mbox{\sf #1}}
\renewcommand{\phi}{\varphi}
\newcommand{\imps}{\mbox{\sc imps}}
\newcommand{\lutins}{\mbox{\sc lutins}}
\newcommand{\mmt}{\mbox{\sc mmt}}

\begin{document}

\title{\bf A New Style of Proof for Mathematics Organized as a Network of
  Axiomatic Theories\thanks{This research was supported by NSERC.}}

\author{
William M. Farmer\thanks{\texttt{wmfarmer@mcmaster.ca.}}\\[2ex]
Department of Computing and Software\\
McMaster University\\
Hamilton, Ontario, Canada
}

\date{30 November 2018}

\maketitle

\begin{abstract}
A \emph{theory graph} is a network of \emph{axiomatic theories}
connected with meaning-preserving mappings called \emph{theory
  morphisms}.  Theory graphs are well suited for organizing large
bodies of mathematical knowledge.  Traditional and formal proofs do
not adequately fulfill all the purposes that mathematical proofs have,
and they do not exploit the structure inherent in a theory graph.  We
propose a new style of proof that fulfills the principal purposes of a
mathematical proof as well as capitalizes on the connections provided
by the theory morphisms in a theory graph.  This new style of proof
combines the strengths of traditional proofs with the strengths of
formal proofs.
\end{abstract}

\input{text}

\bibliography{new-proof-style-for-math} 
\bibliographystyle{plain}

\end{document}

%% file: text.tex
\section{Introduction}\label{sec:introduction}

An \emph{axiomatic theory} (\emph{theory} for short) is a set of
axioms that specifies a set of mathematical structures.  For example,
the usual axioms for a group specify the mathematical structures that
contain an associative binary function with an identity element and an
inverse operation.  Another example is the axioms of a complete
ordered field that uniquely specify the real numbers (up to
isomorphism).

Theories serve as modular units of mathematical knowledge.  Theories
can be constructed by combining smaller theories.  For example, a
theory of fields is a combination of two copies of a theory of groups.
Theories can also be connected to each other via meaning-preserving
mappings called \emph{theory morphisms}\footnote{Theory morphisms are
  also known as \emph{immersions}, \emph{realizations}, \emph{theory
    interpretations}, \emph{translations}, and \emph{views}.}.
(See~\cite{Farmer93} for some illustrative examples of theory
morphisms.)  A theory morphism from a theory $T_1$ to a theory $T_2$
maps the formulas valid in $T_1$ to formulas valid in $T_2$.  For
example, there are two natural theory morphisms from a theory of
groups to a theory of fields.  Theory morphisms serve as information
conduits that enable theory components such as definitions and
theorems to be transported from an abstract theory to a more concrete
theory or equally abstract theory~\cite{BarwiseSeligman97}.

A network of theories connected with theory morphisms is called a
\emph{theory graph}.  The theories are the nodes of the graph, while
the theory morphisms are directed edges.  We will argue in the next section
that the architecture of a theory graph is well suited for organizing
bodies of mathematical knowledge.  A theory morphism connects concepts and
facts in one theory with concepts and facts in another theory that
might be formulated very differently from the first theory.  The
theory morphisms thus make explicit part of the great interconnectedness of
mathematical knowledge.

Neither the traditional proofs given in mathematics papers nor the
formal proofs produced by proof assistants (such as
Coq~\cite{Coq8.8.2}, Isabelle~\cite{IsabelleWebSite}, and
Mizar~\cite{MizarWebSite}) fulfill all the purposes that mathematical
proofs have.  Moreover, they also do not exploit the kind of
connections exhibited within a theory graph.  This paper introduces a
new style of mathematical proof that fulfills the principal purposes
of a mathematical proof as well as capitalizes on the connections
provided by the theory morphisms in a theory graph.  This new style of
proof combines the strengths of traditional proofs with the strengths
of formal proofs.

The rest of the paper is organized as follows.
Section~\ref{sec:theory-graphs} gives an introduction to \emph{theory
  graphs}.  Section~\ref{sec:styles} discusses the different
\emph{styles of proof}, focusing in particular on the traditional and
formal styles.  The notion of a \emph{cross check} that compares a new
result with previous results is presented in
section~\ref{sec:cross-checks}.  The \emph{purposes of a mathematical
  proof} and how well traditional and formal proofs fulfill them are
discussed in section~\ref{sec:purposes}.  A \emph{new style of proof}
in the context of a theory graph is introduced in
section~\ref{sec:new-style}.  And the paper ends with some concluding
remarks in section~\ref{sec:conclusion}.

The major contributions of the work presented in this paper are:

\be

  \item We discuss the importance in mathematics of cross checks.

  \item We describe eight purposes of a mathematical proof and compare
    how well traditional and formal proofs fulfill them.

  \item We introduce a new style of mathematical proof in the context
    of theory graphs that fulfills the purposes of a proof better than
    traditional and formal proofs do.

\ee

\section{Theory Graphs}\label{sec:theory-graphs}

A \emph{theory} is a triple $T=(L,\Sigma,\Gamma)$ where $L$ is a
logic, $\Sigma$ is \emph{language} of $L$, and $\Gamma$ is a set of
formulas of $\Sigma$ called the \emph{axioms} of $T$.  A \emph{model}
for $T$ is an interpretation of $\Sigma$ in $L$ that satisfies all the
members of $\Gamma$.  A model of $T$ can be represented by a theory
whose language has a symbol for each value in the model.

Let $T_i = (L_i,\Sigma_i,\Gamma_i)$ be theories for $i=1,2$.  A
\emph{theory morphism from $T_1$ to $T_2$} is a triple $\Phi =
(T_1,T_2,\phi)$ where $\phi$ is a mapping of the expressions in
$\Sigma_1$ to the expressions in $\Sigma_2$ such that, if $A$ is
formula in $\Sigma_1$ that is a logical consequence of $\Gamma_1$ in
$L_1$, then $\phi(A)$ is a formula in $\Sigma_2$ that is a logical
consequence of $\Gamma_2$ in $L_2$.  Roughly speaking, a theory morphism is a
meaning-preserving syntactic mapping from one theory to another.  The
logics $L_1$ and $L_2$ may be different and the languages $\Sigma_1$
and $\Sigma_2$ may involve totally different vocabulary.  $T_1$ and
$T_2$ are called the \emph{source theory} and \emph{target theory} of
$\Phi$, respectively.

An \emph{instance} of $T_1$ is the target theory of any theory morphism whose
source theory is $T_1$.  An \emph{instance} of an expression $E_1$ in
$\Sigma_1$ is $\phi(E_1)$ for any theory morphism of the form
$(T_1,T_2,\phi)$.  An \emph{inclusion} is a theory morphism $\Phi =
(T_1,T_2,\phi)$ where $T_2$ is an extension of $T_1$ and $\phi$ is the
identity mapping.  Theory morphisms can be composed together: The
\emph{composition} of two theory theory morphisms $\Phi_1 = (T_1,T_2,\phi_1)$
and $\Phi_2 = (T_2,T_3,\phi_2)$ is the theory morphism $\Phi_1 \circ \Phi_2 =
(T_1,T_3,\phi_1 \circ \phi_2)$.

A \emph{theory graph}~\cite{Kohlhase14} is a directed graph whose
nodes are theories and whose edges are theory morphisms.  In a theory
graph, mathematical knowledge is distributed over the set of theories
in the graph.  A theory graph provides an advantageous architecture
for a \emph{digital mathematics library (DML)}~\cite{Bouche10} for the
following reasons:

\be

  \item In accordance with the \emph{little theories
    method}~\cite{FarmerEtAl92b}, the development of a mathematical
    topic can be done in the theory in a theory graph that has the
    most convenient underlying logic, the most convenient level of
    abstraction, and the most convenient vocabulary and then concepts
    and facts produced in the development can be transported to many
    other contexts via the theory morphisms in the theory graph.  For
    example, concepts and facts about a group can be developed in a
    theory of an abstract group and then transported to both the
    additive and multiplicative contexts in a theory of fields.  The
    little theories method thus enables a large body of mathematical
    knowledge to be developed with a minimal amount of redundancy.

  \item The results developed in a theory $T_1$ can be shared with
    another theory $T_2$ that has a different underlying logic,
    vocabulary, or axiomatization as long as $T_2$ exhibits the same
    conceptual structure as $T_1$.  This allows the mathematical
    knowledge in a theory graph to be highly distributed.

  \item Two theories $T_1$ and $T_2$ representing different
    developments of the same mathematical topic based on different
    axiomatizations of the topic can be shown to be equivalent by
    producing a theory morphism from $T_1$ to $T_2$ and another from $T_2$ to
    $T_1$.  The set of equivalent theories can be consolidated into a
    single structure called a \emph{realm}~\cite{CaretteEtAl14}.

  \item The theories in a theory graph can be developed independently
    in parallel and then they can be later integrated with each other
    by defining appropriate theory morphisms between them.  This
    allows a theory graph to built by multiple independent teams
    working in parallel.

  \item Concepts and facts in a theory graph that are relevant to the
    development in a theory $T$ can be found by following theory
    morphisms to $T$ backwards to their source theories.  Concepts and
    facts several ``steps'' away from $T$ can be found by following
    compositions of theory morphisms backwards.  The concepts and
    facts found in this way could reside in theories that are quite
    different from $T$ and possibly even unknown to the developers of
    $T$.

\ee

Substantial software support is needed to realize the benefits of a
theory graph.  Although significant progress has been made in
developing software based on and inspired by theory graphs, there is
not yet a full system for developing and organizing a DML as a theory
graph.

The {\imps} theorem proving system~\cite{FarmerEtAl93,FarmerEtAl96}
represents mathematical knowledge as a theory graph.  However, all the
theories in the {\imps} theory library employ the same underlying
logic, {\lutins}~\cite{Farmer90,Farmer93b,Farmer94}, a version of
Church's type theory with undefinedness and partial functions.  In
{\imps}, theory morphisms are used to transport definitions and
theorems from one theory to another.  They are also used to find for
the user relevant theorems that reside outside of the theory in which
the user is working.

{\mmt}~\cite{RabeKohlhase13} is a foundation-independent framework for
representing mathematical knowledge as a theory graph developed
largely by Florian Rabe within the KWARC research
group~\cite{KWARCWebSite} at Friedrich-Alexander University.  It does
not currently include the kind of tools for developing mathematical
knowledge that proof assistants provide, but it is a very major step
towards a software system that can be used to build a
theory-graph-based DML.  Two other notable KWARC projects that build
on the notion of a theory graph are the OAF project~\cite{OAFWebSite}
on integrating formal libraries and the LATIN
project~\cite{LATINWebSite} on formalizing logics and logic
translations.

Theory morphisms are used in other proof assistants, for example,
Isabelle~\cite{Ballarin14} and PVS~\cite{OwreShankar01}.

\section{Styles of Mathematical Proof}\label{sec:styles}

A \emph{proof} is a deductive argument intended to show that a
mathematical statement is a logical consequence of a set of premises.
There are many styles of proof.  Some proofs \emph{describe} a
deduction of the statement from the premises, while other proofs
\emph{prescribe} the steps needed to produce the deduction.  Many
proofs are presented in a \emph{two-column format} where each line in
the left column is an intermediate result in a deduction and the
corresponding line in the right column explains how the result is
obtained.  Some proofs contain \emph{computations} (e.g., numeric or
algebraic simplifications) or \emph{constructions} (e.g., via
straightedge and compass).  Some are fully \emph{constructive} in the
sense that they strictly adhere to the principles of constructive
logic.  \emph{Geometry proofs} are deductions guided by a geometric
drawing.  \emph{Visual proofs} are presented by a series of diagrams
or an animation.

The proofs presented in mathematical books and articles usually
exhibit a particular style that we call the \emph{traditional proof
  style}.  Proofs of this style are arguments written in a stylized
form of natural language with a heavy use of special symbols.  In
traditional proofs the terminology and notation may be ambiguous,
assumptions may be unstated, and the argument may contain logical
gaps.  However, the reader is expected to be able to resolve the
ambiguities, identify the unstated assumptions, and fill in the gaps
in the argument.  The writer --- whose purpose is to serve some
particular community of readers --- has the freedom to express the
argument in whatever manner is deemed most effective.  This includes
exhibiting other styles of proof within the traditional style.

The \emph{formal proof style} is to present a proof as a derivation in
a proof system for a formal logic.  Formal proofs can be interactively
developed and mechanically checked using proof assistants.  This style
of proof is highly constrained by the logic, proof system, and the
fact that every detail must be verified.  On the other hand, there is
a very high level of assurance that the statement proved is indeed a
theorem of the proof system.  Although the traditional proof style
dominates mathematics, the formal proof style is beginning to make
some modest inroads in mathematical practice.  For example, see the
special issue of the Notices of the AMS on formal
proof~\cite{HalesEtAl08}.

\section{Cross Checks}\label{sec:cross-checks}

A proof by itself does not establish that the theorem it proves is
correct since there is always the possibility of error.  Error is even
possible if the proof is machine checked because the proof may be
valid but the theorem may not be correctly stated.  For example, one
can conjecture that a mathematical object has a certain property,
prove the conjecture, and then conclude that the object does indeed
possess the property.  But the property may have been expressed
incorrectly in a way that is not easily noticed.

Since proofs may be incorrect and theorems may be misstated,
mathematicians are usually reluctant to accept a theorem on only the
basis of its proof.  Georg Kreisel has noted in several of his papers,
e.g., in \cite[p.~126]{Kreisel77} and \cite[p.~145]{Kreisel85}, that a
better way to avoid error than carefully checking a proof is to use
\emph{cross checks} to compare the result with known facts.  For
example, the proof can be checked against similarly structured proofs
and the theorem can be compared with consequences of the theorem or
related versions of the theorem that have been independently proved.
Although cross checks are very important, they are rarely written down
and are not considered as part of either a traditional or a formal
proof.

Here are some examples of cross checks:

\be

  \item Let $P$ be a proof of a theorem $A$.  A cross check would be
    to verify that a proof similar to $P$ proves a theorem similar to
    $A$.

  \item Let $A$ be a theorem asserting that each member of a set $S$
    of objects satisfies a certain property $P$.  A commonly employed
    cross check would be to verify independently from the proof of $A$
    that $P$ is satisfied by certain special members of $S$ like the
    empty set, the empty function, constant functions, etc.

  \item Let $A$ be a theorem and $B$ be a statement that is the
    ``dual'' of $A$ in some sense and that is expected to hold if $A$
    holds.  For instance, if $A$ is statement involving universal
    quantification, $B$ could be the dual statement involving
    existential instead of universal quantification.  A cross check
    would be to verify $B$ independently from the proof of $A$.

  \item Let $A$ be a theorem expressed as an algebraic statement.  A
    cross check would be to verify a geometric analog of $A$
    independently of the proof of $A$.

\ee

In the context of a theory graph, there are two main ways of
representing a cross check.  The first is as a tuple
$(P_1,T_1,P_2,T_2)$ where $P_i$ is a proof in a theory $T_i$ for $
i=1,2$ and $P_1$ and $P_2$ have a similar structure.  This cross check
succeeds if the theorems $P_1$ and $P_2$ prove similar theorems and
fails otherwise.

The second is as a tuple $(A_1,T_1,A_2,T_2,\Phi)$ where:

\be

  \item $A_i$ is a theorem of a theory $T_i$ for $ i=1,2$.

  \item $\Phi$ is a theory morphism $(T_1,T_2,\phi)$.

  \item $A_2$ is expected to follow from $\phi(A_1)$ in $T_2$.

\ee

\noindent
$A_2$ could be, for example, a formulation of $A_1$ in a theory $T_2$
that is a more concrete setting than $T_1$ or the dual of $A_1$ under
some notion of duality captured by $\Phi$.  Notice that, if $\Phi$ is
an inclusion, then $A_2$ is actually expected to follow from $A_1$ in
$T_2$.  In this case, $A_2$ could be a special case of $A_1$ or a
corollary of $A_1$.  This cross check succeeds if $A_2$ indeed follows
from $\phi(A_1)$ in $T_2$ and otherwise fails.

A failed cross check could indicate that a mistake has been made or
that something is not adequately understood.  Thus failed cross checks
are valuable because they can lead to finding hidden mistakes and
making new discoveries.  Also, if the proof or statement of a fully
verified theorem with cross checks is ever modified in the future, the
cross checks can be used to discover errors that are introduced by the
modification.

\section{Purposes of a Mathematical Proof}\label{sec:purposes}

Mathematical proofs serve several purposes.  So what are they?
Various purposes have been discussed in the mathematics
literature~\cite{Bell76,CadwalladerOlsker11,deVilliers90,deVilliers99,Hanna00,Hersh93,LofwallHemmi09}.
Michael de Villiers presents in~\cite{deVilliers99} a list of the
following six purposes:

\be

  \item \emph{Verification} (concerned with the truth of a statement).

  \item \emph{Explanation} (providing insight into why it is true). 

  \item \emph{Systematization} (the organisation of various results
    into a deductive system of axioms, major concepts and theorems).

  \item \emph{Discovery} (the discovery or invention of new results).

  \item \emph{Communication} (the transmission of mathematical
    knowledge).  

  \item \emph{Intellectual challenge} (the
    self-realization/fulfillment derived from constructing a proof).

\ee

We claim that mathematical proofs serve eight principal purposes, four
of which are not on de Villiers' list.  For each of the eight, we
describe what the purpose is and compare how well traditional and
formal proofs fulfill the purpose.

\subsection*{Purpose 1: Communication}

The main purpose of a proof given in a textbook or scientific article
is to \emph{communicate} to the reader why a mathematical statement
follows from a set of premises.  Proofs constructed for communication
are used to convey insight and to build intuition.  The highly
flexible style of traditional proofs is usually a much better vehicle
for communication than the highly constrained style of formal proofs.
This is especially true when the writer is more concerned about
high-level ideas than low-level details (that often can be
mechanically checked by computation).  However, formal proofs can be
much more effective at presenting intricate syntactic manipulations
than traditional proofs.  (This purpose combines de Villiers'
\emph{explanation} and \emph{communication} purposes.)

\subsection*{Purpose 2: Certification}

Another important purpose of a proof is to \emph{certify} that a
mathematical statement follows from a set of premises.  Such a proof
serves as a certificate that can be independently checked.  Since a
traditional proof is written for a particular audience, it may not be
easily checked by someone outside of this audience.  Moreover, a
traditional proof may contain mistakes that are not easily noticed by
a reader, even a reader in the intended audience.  In contrast, a
formal proof can be mechanically checked by software alone.  A formal
proof thus offers the highest level of certification.  (This purpose
includes de Villiers' \emph{verification} purpose.)

\subsection*{Purpose 3: Organization} 

Mathematical knowledge is usually \emph{organized} as a deductive
edifice composed of axioms, definitions, theorems, and proofs.  The
proofs are the threads that hold the edifice together.  Any body of
mathematical knowledge built without proofs will almost certainly
contain falsehoods and contradictions that compromise its deductive
structure.  Both traditional and formal proofs are very effective
tools for organizing mathematical knowledge as a deductive structure,
but formal proofs are somewhat better since their correctness can be
machine checked.  (This purpose is the same as de Villiers'
\emph{systematization} purpose.)

\subsection*{Purpose 4: Discovery}

A proof is often formulated to be a provisional argument that a
mathematician can use to \emph{discover} new theorems.  This idea is
brilliantly expressed in \emph{Proofs and Refutations} by Imre
Lakatos~\cite{Lakatos76}.  See also Yehuda Rav, ``Why Do We Prove
Theorems?''~\cite{Rav99}.  Traditional proofs are well suited for
expressing provisional arguments that can be analyzed by humans.
Formal proofs are too rigid to express provisional arguments and thus
are poorly suited for this task.  On the other hand, machines can be
used to discover various kinds of structure embodied in a formal
proof, but it is much more difficult to analyze traditional proofs in
this way.  (This purpose is the same as de Villiers' \emph{discovery}
purpose.)

\subsection*{Purpose 5: Learning}

The most effective way to \emph{learn} mathematics is to read and
write proofs.  Traditional proofs are today generally much easier to
read and write than formal proofs.  However, a reader of a traditional
proof may have to work hard to resolve ambiguities, identify unstated
assumptions, and fill in the gaps in the argument, and a writer may
have to work hard to verify that each step of the argument is valid.
With effective software support, reading and writing formal proofs
could become almost as easy as reading and writing traditional proofs.
(This purpose is not explicitly included in de Villiers' list of
purposes.)

\subsection*{Purpose 6: Universality}

A proof is \emph{universal} if it is expressed without any superfluous
ideas and can thus be applied in every context in which the conditions
of the proof hold.  Universality is not absolute; it depends on
audience and context.  A proof can be universal with respect to one
audience and context but not with respect to another.  Traditional
proofs can be expressed in a universal manner, but the underlying
mathematical foundation is usually implicit.  Traditional proofs are
thus untethered; they do not have a precise mathematical home.  Formal
proofs have a precise mathematical home, but the home is usually not
connected to many other contexts in which the proof can be applied.
Hence both traditional and formal proofs fall short in achieving
universality.  (This purpose is not included in de Villiers' list of
purposes.)

\subsection*{Purpose 7: Coherency}

A theorem is \emph{coherent} with a body of mathematical knowledge if
it properly fits into the body without any contradictions or
unexpected relationships.  A traditional or formal proof by itself
does not establish that the theorem it proves is coherent with other
mathematical knowledge.  Coherency is established by cross checks.
Although cross checks are very important, they are rarely written down
and are not considered as part of either a traditional or a formal
proof.  (This purpose is not explicitly included in de Villiers' list
of purposes.)

\subsection*{Purpose 8: Beauty}

Mathematics is a utilitarian art form like architecture or industrial
design.  The desire to create \emph{beauty} (what mathematicians call
\emph{elegance}) is one of the strongest driving forces in
mathematics.  Mathematicians seek to develop proofs that are beautiful
as well as correct.  Indeed some mathematicians will not accept a
theorem until an elegant proof of the theorem has been found.  It is
safe to say that most mathematicians find it easier to write beautiful
proofs in the highly flexible traditional proof style than in the
highly constrained formal proof style.  (This purpose is not included
in de Villiers' list of purposes.)

\subsection*{Summary}

Table~\ref{tab:summary} summarizes the differences between traditional
and formal proofs.  As can be seen, neither traditional proofs nor
formal proofs fulfill the eight purposes that we claim mathematical
proofs have.  Furthermore, both styles lack the capacity to fully
achieve universality and coherency.

\begin{table}[ht]
\bc
\begin{tabular}{|l|c|c|}
\hline
& \textbf{Traditional Proofs} &\textbf{Formal Proofs}\\
\hline
\textbf{Communication} &
{\high} &
{\medlow}\\
\hline
\textbf{Certification} & 
{\medlow} &
{\high}\\
\hline
\textbf{Organization} & 
{\medhigh} &
{\high}\\
\hline
\textbf{Discovery (Human)} & 
{\high} &
{\low}\\
\hline
\textbf{Discovery (Machine)} & 
{\low} &
{\high}\\
\hline
\textbf{Learning (Reading)} &
{\medhigh} &
{\medlow}\\
\hline
\textbf{Learning (Writing)} &
{\medhigh} &
{\medlow}\\
\hline
\textbf{Universality} & 
{\medlow} &
{\medlow}\\
\hline
\textbf{Coherency} & 
{\low} &
{\low}\\
\hline
\textbf{Beauty} & 
{\high} &
{\low}\\
\hline
\end{tabular}

\medskip

$\high$ : high; $\medhigh$ : medium high; $\medlow$ : medium low; $\low$ : low.
\ec
\caption{Traditional vs.~Formal Proofs}
\label{tab:summary}
\end{table}

\section{A New Style of Proof}\label{sec:new-style}

Since traditional and formal proofs do not adequately achieve
universality and coherency, they are not adequate for building theory
graphs.  We therefore propose a new style of proof that fulfills
universality and coherency as well as other six purposes described in
the previous section.  Let \mname{TG} be a theory graph.  A proof in
\mname{TG} of this new proof style has four components:

\be

  \item A \emph{home theory} $\mname{HT} = (\mname{Log}, \mname{Lang},
    \mname{Axms})$ where {Log} is a formal logic, \mname{Lang} is a
    language in \mname{Log}, and \mname{Axms} is a set of formulas in
    \mname{Lang}.

  \item A \emph{theorem} \mname{Thm} that is a formula in \mname{Lang}
    purported to be a logical consequence of \mname{Axms}.

  \item An \emph{argument} \mname{Arg} that shows \mname{Thm} is a
    logical consequence of \mname{Axms}.

  \item A set \mname{CC} of \emph{cross checks} of the two forms
    mentioned in section~\ref{sec:cross-checks} that compare
    \mname{Arg} with similar arguments in \mname{TG} and \mname{Thm}
    with related theorems in \mname{HT} or in other theories in
    \mname{TG}.

\ee

The home theory \mname{HT} is a node in \mname{TG} and a formal
context for the proof.  It is connected via theory morphisms to other
theories in \mname{TG}.  Ideally, the home theory is at the optimal
level of abstraction for the proof and contains only the concepts and
assumptions needed to express the proof's argument and theorem.

The theorem \mname{Thm} is a formal statement of what the proof's
argument shows.  It can be transported via appropriate theory morphisms to
other theories in which the conditions of the proof hold.  \mname{HT}
and \mname{Thm} together thus serve as a specification of the set of
theories $T$ and formulas $A$ in \mname{TG} such that $T$ is an
instance of \mname{HT} and $A$ is an instance of \mname{Thm} under
some theory morphism.  In this way, the proof fulfills the purpose of
universality.

The argument \mname{Arg} has both a traditional component for
communication, organization, human-oriented discovery, learning, and
beauty and a formal component for certification, organization,
learning, and machine-oriented discovery.  The two components are
tightly integrated so that, for example, a reader of the traditional
component can switch, if desired, to the formal component when a gap
in the argument is reached.  It is not necessary that the formal
component is a complete formal proof of the theorem.  The formal
component can even be totally absent.  Thus the proof is
\emph{flexiformal}~\cite{Kohlhase13} in the sense that it is mixture
of formal and informal components.

The set of cross checks should be carefully chosen to show that the
theorem is coherent with the web of previously established facts in
\mname{TG}.  With the set \mname{CC} the proof thus fulfills the
purpose of coherency.

In summary, the new style of proof we propose is a mixture of the
traditional and formal proof styles in which the context of the proof
and the statement proved are formal, the argument of the proof is
expressed in a traditional style, and parts of the argument may be
integrated with formal derivations.  The home theory of the proof is a
node in a theory graph of a \mname{TG} that is an optimal expression
of the context of the proof.  And the cross checks of the proof
connect the proof and the theorem to similar proofs and related
theorems in the theory graph.

\section{Conclusion}\label{sec:conclusion}

We have shown that a theory graph is a network of theories connected
by theory morphisms in which mathematical knowledge is distributed
across the network of theories.  The underlying logics of the theories
can be different and the languages of the theories can vary greatly.
The theories can organized according to the little theories method and
can be developed independently in parallel.  The theory morphisms
capture many of the connections in the mathematical knowledge
represented in the theory graph.  As a result, they can be used to
find concepts and facts relevant to a theory $T$ that reside outside
of $T$ in other, possibly quite different, theories.

With these attributes, a theory graph is well suited to be the
architecture for a large-scale, multifoundational, highly connected,
and highly distributed DML.  This is particularly true for a DML whose
mathematical knowledge is intended to be formal or flexiformal.  The
obvious example of such a DML is the \emph{Global Digital Mathematics
  Library (GDML)}~\cite{IonWatt17} proposed by the International
Mathematical Union (IMU)~\cite{IMUWebSite}.

Proofs have a crucial role to play in building a GDML.  However,
traditional and formal proofs do not adequately fulfill all the
purposes of a proof that we presented in section~\ref{sec:purposes}.
Traditional proofs are good for communication, organization, human
discovery, learning, and beauty, while formal proofs are good for
certification, organization machine discovery.  But neither
traditional nor formal proofs are especially good for universality and
coherency.

To capitalize on the structure offered by a theory graph, we have
proposed a new style of proof that merges the traditional and formal
styles of proof, achieves universality using the little theories
method, and incorporates cross checks to establish coherency.  We
believe this proof style will promote the development of highly
structured DMLs while preserving the benefits of both traditional and
formal proofs.

\section*{Acknowledgments}

The author would like to thank the referees for their comments. This
research was supported by NSERC.